\documentclass[12pt]{article}
\usepackage{amsfonts,amssymb,epsf,cite}
\begin{document}
\title{Quintessence and black holes}
\author{V.V.Kiselev\\[0mm] 
\small 
\sl State Research Center "Institute for High Energy Physics" \\[-2mm]
\small {\sl Protvino, Moscow region, 142280 Russia}\\[-1.5mm]
\small \sl Fax: +7-0967-744739,
E-mail: {kiselev@th1.ihep.su}
}
\date{}
\maketitle
\begin{abstract}
We present new static spherically-symmetric exact solutions of Einstein
equations with the quintessential matter surrounding a black hole charged or
not as well as for the case without the black hole. A condition of 
additivity and linearity in the energy-momentum tensor is introduced, which
allows one to get correct limits to the known solutions for the electromagnetic
static field implying the relativistic relation between the energy density and
pressure, as well as for the extraordinary case of cosmological constant, i.e.
de Sitter space. We classify the horizons, which evidently reveal themselves in
the static coordinates, and derive the Gibbons-Hawking temperatures. An example
of quintessence with the state parameter $w=-2/3$ is discussed in detail.
\end{abstract}

\section{Introduction}
Recent astronomical observations convincingly show the accelerating expansion
of Universe \cite{Perl}, implying a valuable presence of state with the
negative pressure. The origin of the negative pressure could be twofold. The
first is the cosmological constant, while the second is the so-called
quintessence with the state equation given by the relation between the pressure
$p_q$ and the energy density $\rho_q$, so that $p_q = w_q\,\rho_q$ at $w_q$ in
the range of $-1 < w_q < -\frac{1}{3}$, which causes the
acceleration\footnote{The phenomenological constraints on the value of $w_q$
are given in \cite{Stein-Chiba}.}. The border case of $w_q=-1$ of the
extraordinary quintessence covers the cosmological constant term. As was
recognized \cite{Suss,B,W}, the acceleration makes a challenge for the
consistent theory of quantum gravity because of the outer horizon, which does
not allow one to introduce the observable S-matrix in terms of asymptotic past
and future states. The outer horizon of de Sitter space significantly
differs from the inner horizon of black hole, which has asymptotically flat
space far away from the black hole. 

The future horizon for the Robertson--Walker metric with the accelerating scale
factor caused by the quintessence was studied in \cite{Suss}. In this paper we
investigate the Einstein equations for the static spherically-symmetric
quintessence surrounding a black hole and solve them exactly for a specific
choice of a free parameter, characterizing the energy-momentum tensor of the
quintessence, ${T_{\mu}}^\nu$. So, under the spherical symmetry of a static
state one can write down a general expression for the time and spatial
components
$$
{T_t}^t = A(r), \qquad {T_t}^j = 0,\qquad
{T_i}^j = C(r)\,r_i\,r^j +B(r)\,{\delta_i}^j,
$$
with the metric $${\rm d}s^2 = g_{tt}(r)\,{\rm d}t^2-g_{rr}(r)\,{\rm d}r^2-
r^2({\rm d}\theta^2+\sin^2\theta\,{\rm d}\phi^2),$$
and after averaging over the angles of isotropic state we get
$$
\langle {T_i}^j\rangle = D(r)\,{\delta_i}^j, \qquad
D(r) = -\frac{1}{3}\,C(r)\,r^2 +B(r).
$$
For the quintessence we have
$$
D(r) = - w_q\, A(r).
$$
Therefore, fixing the state parameter $w_q$ gives the expression for the
function $D(r)$ being the combination of $C(r)$ and $B(r)$, in terms of density
$A(r)$, while $w_q$ itself does not provide us with the complete information on
the form of energy-momentum tensor in the static spherically-symmetric case. 

The problem with the free quintessence under the condition of 
$$
C(r) \equiv 0
$$
was considered in \cite{Ch} and in \cite{GD} by P.F.Gonzalez-Diaz. The result
was the metric, which has no horizon, no `hair' and no black hole, i.e. the
metric does not allow one to `embrace a black hole term'.

In the present paper we consider nonzero $C(r)$ proportional to $B(r)$, so that
the exact solutions with the black hole charged or not are possible, including
the generalization to the asymptotically flat or de Sitter space. The
appropriate constant coefficient $C(r)/B(r)$ is defined by the condition of
additivity and linearity, and it allows us to get correct limits for the well
know cases of the relativistic matter of electric field for the charged black
hole ($w_q =\frac{1}{3}$), the dust matter of $w_q=0$, and the extraordinary
quintessence with $w_q=-1$, i.e. the cosmological constant.

We find that in the static coordinates the exact solutions of
spherically-symmetric Einstein equations for the quintessence produce the
outer horizons at $-1 < w_q < -\frac{1}{3}$ in the cases of free state as well
as for the black hole surrounded by the quintessence.

In section \ref{1} we construct a general solution of spherically-symmetric 
static Einstein equations for the quintessence satisfying the condition of the
additivity and linearity, which allows us to cover the problems with the black
hole charged or not in the flat or de Sitter space. Section \ref{2} is devoted
to the consideration of examples and limits known for the relativistic matter
and the cosmological constant. The Gibbons--Hawking temperatures \cite{GH} of
the free quintessence as well as for the black hole surrounded by the
quintessence with $w_q=-\frac{2}{3}$ are described in Section \ref{3}. The
results are summarized in Conclusion.

\section{Exact solutions in static coordinates\label{1}}
Following the notations in \cite{Landafshiz}, we parameterize the interval of
spherically-symmetric static gravitational field by
\begin{equation}
{\rm d}s^2 = e^{\nu}\,{\rm d}t^2-e^{\lambda}\,{\rm d}r^2 - r^2({\rm
d}\theta^2+\sin^2\theta\,{\rm d}\phi^2),
\end{equation}
with the functions $\nu =\nu(r)$ and $\lambda =\lambda(r)$. Then, in units with
the normalization of gravitational constant $G$ by $4\pi G =1$, the Einstein
equations get the form
\begin{eqnarray}
&&2\,{T_t}^t = - e^{-\lambda}\left(\frac{1}{r^2} -
\frac{\lambda^\prime}{r}\right) + \frac{1}{r^2}, \\[2mm]
&&2\,{T_r}^r = - e^{-\lambda}\left(\frac{1}{r^2}
+ \frac{\nu^\prime}{r}\right) + \frac{1}{r^2}, \\[2mm]
&&2\,{T_\theta}^\theta = 2\,{T_\phi}^\phi = - \frac{1}{2}
e^{-\lambda}\left(\nu^{\prime\prime}+
\frac{{\nu^\prime}^2}{2} + \frac{\nu^\prime-\lambda^\prime}{r} -
\frac{\nu^\prime\,\lambda^\prime}{2}\right), 
\end{eqnarray}
with $u^\prime \stackrel{\mbox{\tiny def}}{=} {\rm d}u(r)/{\rm d}r$. The
appropriate general expression for the energy-momentum tensor of quintessence
is given by
\begin{eqnarray}
&&{T_t}^t = \rho_q(r), \\[2mm]
&&{T_i}^j  =  \rho_q(r)\,\alpha
\left[-(1+3\,B)\frac{r_i\,r^j}{r_n r^n}+B\,{\delta_i}^j\right],
\end{eqnarray}
so that the spatial part is proportional to the time component with the
arbitrary parameter $B$ depending on the internal structure of quintessence.
The isotropic averaging over the angeles results in 
\begin{equation}
\left\langle {T_i}^j\right\rangle = - \rho_q(r)\,\frac{\alpha}{3}\,
{\delta_i}^j = - p_q(r)\,{\delta_i}^j,
\end{equation}
since we put $\langle r_i\,r^j\rangle = \frac{1}{3}\, {\delta_i}^j\,r_n r^n$.
Therefore, we derive the relations
\begin{equation}
p_q = w_q\, \rho_q, \qquad w_q = \frac{1}{3}\,\alpha.
\end{equation}
The quintessential state has
$$
-1< w_q < 0 \quad \Longrightarrow \quad -3 < \alpha < 0.
$$
Let us define a principle of additivity and linearity by the equality, putting
the relation between the metric components,
\begin{equation}
{T_t}^t = {T_r}^r \quad \Longrightarrow \quad \lambda + \nu = 0,
\end{equation}
without any lose of generality due to the static coordinate system fixed by
the above gauge of $\lambda + \nu = \mbox{const} = 0$, since the constant can
be nullified by an appropriate rescaling of time. 

Then, substituting for
$$
\lambda = -\ln(1+f),
$$
we get linear differential equations in $f$, so that 
\begin{eqnarray}
&&{T_t}^t  =  {T_r}^r = - \frac{1}{2\,r^2}\, (f + r\,f^\prime), 
\label{t}\\[2mm]
&&{T_\theta}^\theta  =  {T_\phi}^\phi = - \frac{1}{4\,r}\,(2 f^\prime
+r\,f^{\prime\prime}).
\label{th}
\end{eqnarray}
Eqs. (\ref{t}) and (\ref{th}) imply that the linear sum of various solutions
for $f$ is transformed to the sum of corresponding terms in the energy-momentum
tensor of matter,
$$
\sum_{n} c_n\, f_n(r) \quad \Longrightarrow \quad \sum_{n} c_n
{T_\mu}^\nu[f_n(r)].
$$
The condition of the additivity and linearity fixes the free parameter of the
energy-momentum tensor for the matter
\begin{equation}
B = -\frac{3 w_q+1}{6 w_q},
\end{equation}
which implies
\begin{eqnarray}
&&{T_t}^t  =  {T_r}^r = \rho_q,
\label{tq} \\[2mm]
&&{T_\theta}^\theta  =  {T_\phi}^\phi = -\frac{1}{2}\,\rho_q\,(3 w_q+1).
\label{thq}
\end{eqnarray}
Making use of (\ref{t})-(\ref{th}) with (\ref{tq})-(\ref{thq}), we deduce
$$
(3 w_q+1)\, f + 3 (1+w_q)\,r\, f^\prime + r^2\,f^{\prime\prime} = 0,
$$
with two solutions in the form
\begin{eqnarray}
&&~~f_q  =  \frac{c}{r^{3 w_q+1}}, \\[2mm]
&&f_{\rm BH}  =  -\frac{r_g}{r},
\end{eqnarray}
where $c$ and $r_g$ are the normalization factors. The function of $f_{\rm BH}$
represents the ordinary Schwarzschild solution for the point-like black hole,
and it coincides with the particular choice of dust matter with $w_q=0$ in
$f_q$, which gives $\rho_q =0$ at $r\neq 0$.

If we put the density of energy to be positive, $\rho_q > 0$, then from the
formula
$$
\rho_q = \frac{c}{2}\,\frac{3 w_q}{r^{3(1+w_q)}}
$$
we deduce that the sign of the normalization constant should coincide with the
sign of the matter state parameter, i.e.
$$
c\,w_q \geqslant 0,
$$
implying that $c$ is negative for the quintessence.

The curvature has the form
\begin{equation}
R = 2\,{T_\mu}^\mu = 3\,c\,w_q\,\frac{1-3w_q}{r^{3(w_q+1)}},
\end{equation}
and it has the singularity at $r=0$, if $w_q\neq \{0,\frac{1}{3},-1\}$.

Thus, we have found a general form of exact spherically-symmetric solutions for
the Einstein equations describing black holes surrounded by the quintessential
matter with the energy-momentum tensor, which satisfies the condition of the
additivity and linearity in accordance with eqs.(\ref{tq})--(\ref{thq}), so
that the metric is given by
\begin{equation}
{\rm d}s^2 = \left[1-\frac{r_g}{r}-\sum_n \left( \frac{r_n}{r} \right)^{3w_n+1}
\right]\,{\rm d}t^2-\frac{{\rm d}r^2}{\displaystyle\left[1-\frac{r_g}{r}-\sum_n
\left(\frac{r_n}{r} \right)^{3w_n+1} \right]} - r^2({\rm d}\theta^2 -
\sin^2\theta\, {\rm d}\phi^2),
\label{general solution}
\end{equation}
where $r_g = 2 M$, $M$ is the black hole mass, $r_n$ are the dimensional
normalization constants, and $w_n$ are the quintessential state parameters.
\section{Some examples\label{2}}
Let us start with checking the important limits, i.e. the charged or
neutral black holes in the asymptotically flat and de Sitter spaces.

The electrically charged black hole surrounded by the static
spherically-symmetric electric field corresponds to the case with the
relativistic matter state parameter $w=\frac{1}{3}$. The general solution of
(\ref{general solution}) with
$$
g_{rr} = -\frac{1}{g_{tt}}
$$
gives the Reissner--Nordstr\"om metric for the charged black hole or the
Schwarzschild one at the charge equal to zero, $e=0$, so that
\begin{equation}
g_{tt} = 1-\frac{r_g}{r}+\frac{e^2}{r^2},
\end{equation}
as well as the transparent generalization to the case of de Sitter space
\begin{equation}
g_{tt}^{\rm dS} = 1-\frac{r_g}{r}+\frac{e^2}{r^2}-\frac{r^2}{a^2}.
\end{equation}
Therefore, we get convincing reasons to see that the condition of the
additivity and linearity allows us to maintain the valid limits to the known
solutions for the spherically-symmetric black holes, which have a rigorous
physical meaning.

The solution for the Reissner--Nordstr\"om--de Sitter black hole surrounded by
the quintessence gives
\begin{equation}
g_{tt}^{\rm QdS} = 1-\frac{r_g}{r}+\frac{e^2}{r^2}-\frac{r^2}{a^2}-\left(
\frac{r_q}{r} \right)^{3w_q+1},
\label{gen}
\end{equation}
which has the meaningful limits with no charge ($e=0$), no de Sitter curvature
($a^2\to \infty$) as well as the self-gravitating quintessence without the
black hole ($r_g=0$ and $e=0$).

In what follows, the characteristic example we are going to study in detail
corresponds to the choice of $w_q=-\frac{2}{3}$, so that
\begin{equation}
\tilde g_{tt}^{\rm QdS} = 1-\frac{r_g}{r}+\frac{e^2}{r^2}-\frac{r^2}{a^2}-
\frac{r}{r_q},
\label{ex}
\end{equation}
whereas the free quintessence determines the interval
\begin{equation}
{\rm d}\tilde s^2 = \left[1-\frac{r}{r_q} \right]\,{\rm
d}t^2-\frac{1}{\displaystyle 1-\frac{r}{r_q}}\,{\rm d}r^2 - r^2({\rm
d}\theta^2+\sin^2\theta\,{\rm d}\phi^2),
\end{equation}
which we compare with the case of de Sitter metric
\begin{equation}
{\rm d}\bar s^2 = \left[1-\frac{r^2}{a^2} \right]\,{\rm
d}t^2-\frac{1}{\displaystyle 1-\frac{r^2}{a^2}}\,{\rm d}r^2 - r^2({\rm
d}\theta^2+\sin^2\theta\,{\rm d}\phi^2).
\end{equation}
The curvature of this static quintessential state is equal to
$$
\tilde R_{\rm Q} = \frac{6}{r_q\,r}.
$$
Under the coordinate transformation by
$$
\chi =\frac{1}{2r_q}\int_0^r \frac{{\rm d}r}{\tilde g_{tt}(r)} =
-\frac{1}{2}\ln\left(1-\frac{r}{r_q}\right),
$$
and $t\to t/2r_q$, we find the hyperbolic representation of interval
\begin{equation}
{\rm d}\tilde s^2 = 4r_q^2\,e^{-2\chi}\,\left[{\rm d}t^2 - {\rm d}\chi^2 -
\sinh^2\chi\,({\rm d}\theta^2+\sin^2\theta\,{\rm d}\phi^2)\right],
\label{dStilde}
\end{equation}
with $0 \leqslant \chi < \infty$, while an appropriate similar transformation
of de Sitter metric gives
\begin{equation}
{\rm d}\bar s^2 = \frac{a^2}{\cosh^2\chi}\,\left[{\rm d}t^2 - {\rm d}\chi^2 -
\sinh^2\chi\,({\rm d}\theta^2+\sin^2\theta\,{\rm d}\phi^2)\right].
\label{dSbar}
\end{equation}
Eqs. (\ref{dStilde}) and (\ref{dSbar}) reveal that at $\chi\to\infty$, i.e.
near the horizon, these two metrics are infinitely close to each other up to a
constant conformal factor. The curvature takes the form
$$
\tilde R_{\rm Q} = \frac{3}{r_q^2}\,\left(1+\frac{1}{\tanh\chi}\right).
$$
Next, under the Fock transformation from the hyperbolic coordinates to the
flat ones,
\begin{eqnarray}
&&\tau =e^t\,\cosh \chi,\\[2mm]
&&\rho =e^t\,\sinh \chi,
\end{eqnarray}
we find
\begin{equation}
{\rm d}\tilde s^2 = 4r_q^2\,\frac{1}{(\tau+\rho)^2}\,\left[{\rm d}\tau^2 - {\rm
d}\rho^2 -
\rho^2\,({\rm d}\theta^2+\sin^2\theta\,{\rm d}\phi^2)\right],
\label{dStildef}
\end{equation}
with the curvature
$$
\tilde R_{\rm Q} = \frac{3}{r_q^2}\,\frac{\tau+\rho}{\rho}.
$$
The conformal-time expression for the metric of de Sitter is given by 
\begin{equation}
{\rm d}\bar s^2 = a^2\,\frac{1}{\tau^2}\,\left[{\rm d}\tau^2 - {\rm
d}\rho^2 - \rho^2\,({\rm d}\theta^2+\sin^2\theta\,{\rm d}\phi^2)\right],
\label{dSbarf}
\end{equation}
which allows the treatment in terms of Friedmann--Robertson--Walker evolution
of flat universe, in contrast to (\ref{dStildef}), which reveals the dependence
of conformal factor on both the time and the distance from the singular point
at $\rho=0$.

\begin{figure}[th]
\hspace*{3.7cm}
\epsfxsize=10cm \epsfbox{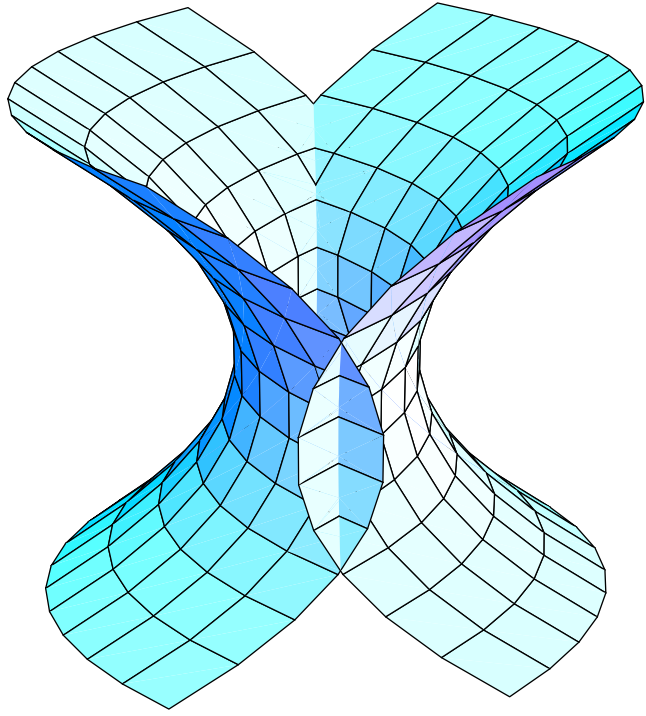}
\caption{The surface of (\ref{sur}) at $z_2=z_3=0$.}
\label{surf}
\end{figure}

The constant curvature of de Sitter space allows one to write down the
invariant presentation in the 5-dimensional Minkowski space\footnote{The metric
is given by diag$(1,-1,-1,-1,-1)$.} in terms of
\begin{eqnarray}
&&\bar z_0 = \sqrt{a^2-r^2} \sinh [t/a],\nonumber\\[2mm]
&&\bar z_4 = \sqrt{a^2-r^2} \cosh [t/a],\\[2mm]
&&\bar z_i = r_i, \qquad i=\{1,2,3\},\nonumber
\end{eqnarray}
so that
$$
\bar z_0^2 - \bar z_4^2 - \bar z_i^2 = -a^2,
$$
with the SO(1,4) invariance. For the static quintessential state one can
introduce the analogous variables in the 5-dimensional Minkowski space 
\begin{eqnarray}
&&\tilde z_0 = 2r_q\,\sqrt{1-\frac{r}{r_q}} \sinh
\left[\frac{t}{2r_q}+\sqrt{1-\frac{r}{r_q}}\right],\nonumber\\[2mm]
&&\tilde z_4 = 2r_q\,\sqrt{1-\frac{r}{r_q}} \cosh
\left[\frac{t}{2r_q}+\sqrt{1-\frac{r}{r_q}}\right],\\[2mm]
&&\tilde z_i = r_i, \qquad i=\{1,2,3\}, \nonumber
\end{eqnarray}
so that
\begin{equation}
\tilde z_0^2 - \tilde z_4^2 - \tilde z_i^2 = -\left(2r_q-\sqrt{\tilde
z_i^2}\right)^2,
\label{sur}
\end{equation}
with the evident 4-parametric symmetry of SO(1,1)$\otimes$SO(3).
The corresponding surface is not smooth. It is represented in Fig. \ref{surf}
by putting $z_2=z_3=0$.
The intersections of two sheets take place at $z_1=0$, i.e. at $r=0$, where
the curvature has the singular point, and the continuation into the external
wings corresponds to the negative $r$ and, hence, the negative curvature.
\section{Horizons and thermodynamics\label{3}}
Using (\ref{gen}) we see that the free quintessence generates the outer
horizon of de Sitter kind at $r=r_q$ if
$$
-1 < w_q < - \frac{1}{3},
$$
or the inner horizon of the black hole kind at $r=r_q$ if
$$
- \frac{1}{3} < w_q < 0.
$$
The Gibbons--Hawking temperature is given by the equation
\begin{equation}
{\rm T} = \frac{1}{4\pi}\left|\frac{{\rm d} g_{tt}(r_q)}{{\rm d} r}\right|,
\label{GH}
\end{equation}
which results in 
\begin{equation}
{\rm T}_{\rm Q} = \frac{1}{4\pi}\,\frac{|3w_q+1|}{r_q}.
\label{GHQ}
\end{equation}
In the characteristic case of the neutral black hole surrounded by the
quintessence with \mbox{$w_q=-\frac{2}{3}$} in (\ref{ex}) we get the inner and
outer
horizons,
\begin{eqnarray}
&& r_{\rm in} = \frac{1}{2}\,\left(r_q - \sqrt{r_q^2-4 r_q r_g}\right),\\[2mm]
&& r_{\rm out} = \frac{1}{2}\,\left(r_q + \sqrt{r_q^2-4 r_q r_g}\right),
\end{eqnarray}
if 
$$
r_q > 4r_g,
$$
whereas
$$
r_g < r_{\rm in} < r_{\rm out} < r_q.
$$
Then, the temperatures at the horizons are given by the following expression:
\begin{equation}
{\rm T}_{\rm QBH}^{\rm in,\, out} = \frac{1}{4\pi}\,\left(\frac{r_g}{r^2_{\rm
in,\,out}}-\frac{1}{r_q}\right),
\end{equation}
so that we observe the hierarchy
$$
{\rm T}_{\rm Q} < {\rm T}_{\rm QBH}^{\rm out} < {\rm T}_{\rm QBH}^{\rm in} <
{\rm T}_{g},
$$
implying that the quintessence decreases the temperature of isolated black hole
${\rm T}_{g}$, while the black hole increases the temperature of free
quintessence. 

The minimal temperature ${\rm T}_{\rm QBH}^{\rm in,\, out}=0$ corresponds to
the degenerate case of $r_q = 4r_g$.
\section{Conclusion\label{4}}

We have got exact solutions of Einstein equations for the static
spherically-symmetric quintessential state free or surrounding a black hole
charged or not. These solutions are possible under the special choice of
internal parameter in the energy-momentum tensor of quintessence, depending on
the state parameter $w_q$. The corresponding condition of the additivity and
linearity results in the linear equations for the various matter terms
consistent with the simple summation of their contributions in the total
energy-momentum tensor.

The free spherically-symmetric quintessential state in the static coordinates
reveals the outer horizons of de Sitter kind at $w_q<-\frac{1}{3}$ in
agreement with \cite{Suss}. We have calculated also the temperatures at
horizons in some specific examples.

To the moment we have presented the problem with the quintessence itself being
a perfect fluid with no physical motivations, so that the consideration has
followed a general model-independent way. In practice, however, one usually
handles a field-theoretical approach in terms of Lagrangian formalism. So, the
quintessence is treated as a scalar field with a specific potential and,
probably, a kinetic term ($k$-sense). Spherically symmetric solutions in
dimensionally reduced spacetimes were in detail investigated by Mignemi and
Wiltshire in  \cite{NW1}. In that study the vacuum Einstein equations in
$2+m+n$ dimensions $=\mbox{time}+\mbox{radius}+m\mbox{-dimensional
sphere}+n\mbox{-dimensional compact extra-space}$ are equivalent to the field
equations for a scalar field coupled to the gravity in $2+m$ dimensions.
This field has an exponential potential, which generally leads to the perfect
fluid approximation with a variable state parameter. Such the approximation is
meaningful in the cosmological models of Universe evolution. The authors of
\cite{NW1} found that the no-hair theorem remains valid, while some non-trivial
solutions (not Schwarzschild ones) can be possible under non-flat asymptotics.
The critical points leading to naked singularities were described. The
consideration were further extended to the black holes in higher-derivative
gravity theories by \cite{NW2}, wherein the authors classified all static
spherically symmetric solutions of D-dimensional gravity coupled to a scalar
field with a potential consisting of a finite sum of exponential terms. The
analysis for global properties of static spherically symmetric charged dilaton
spacetimes with a Liouville potential was done by Poletti and Wiltshire in
\cite{PW}. A treatment of scalar field in the context of dark matter and dark
energy was developed in \cite{G1,G2}. So, the static solutions for the
spherically symmetric scalar field were found for the description of flat
rotation curves in spiral galaxies provided the potential of the field is
exponential, while the energy density and presure behaves as $1/r^2$ with the
distance. The properties of solutions are in agreement with the general
analysis in \cite{NW1,NW2,PW}. The problem of such the treatment is twofold.
First, the potential has a negative sign in contrast to the cosmological
applications. Second, small rotation velocities lead to the extra compact
dimension $n\ll 1$. Nevertheless, authors of \cite{G3} found that the origin of
quintessence at the cosmological and galaxy scales could be significantly
different. Indeed, the flat rotation curves in the geodetic analysis make a
preference for the quintessence with the state parameter $w=-\frac{1}{3}$ in
contrast to $w$ close to $-1$ for the accelerating evolution of Universe. So,
the exponential potentials as well as the scalar fields responsible for the
dark matter and dark energy can be different.

Next, the $1/r^2$-falling of energy-momentum tensor for the scalar field
responsible for the flat rotation curves \cite{G1} can be obtained in the
general solution presented in this work by implementing the limit $w_q\to
-\frac{1}{3}$ yielding 
$$
{T_t}^t = {T_r}^r \sim \frac{1}{r^2},\qquad {T_\theta}^\theta =
{T_\phi}^\phi=0,
$$
while in \cite{G1} one found
$$
\left|{T_t}^t\right|\sim \left|{T_\theta}^\theta\right| =
\left|{T_\phi}^\phi\right| \ll \left|{T_r}^r
\right| \sim \frac{1}{r^2},
$$
in the limit of scalar field alone. We see that the physical contents of these
two approaches can be similar if we suggest that the perfect fluid in the
present paper is represented by a coherent state formed by a scalar field and
a cold dark matter (the pressure equals zero). Of course, such the suggestion
should be tested in a model way elsewhere. 

Further, the angle surplus caused by the scalar field as found in \cite{G1} can
be obtained in the perfect fluid approach used, too. Indeed, the limit of
$w_q\to -\frac{1}{3}$ implies that the metric component $g_{rr}$ tends to a
constant value. Therefore, we can rescale the definition of radius by $r\to
r/\sqrt{-g_{rr}}$, so that the sphere surface gets the angle defect (surplus or
deficit) since its area becomes $4\pi r^2/g_{rr}$ instead of $4\pi r^2$. In
addition, an interesting point is that in the case $w_q =-\frac{2}{3}$, which
we have concerned with in section \ref{2}, 
in the absence of a de Sitter term
the spacetime is not asymptotically flat but has an asymptotic defect angle,
just as was noted for the model in ref. \cite{G1}. Note, that in typical
quintessence models the cosmological constant is omitted since the quintessence
field is taken play the role of driving the acceleration of the universe at
late times. If that is the case, then presumably the asymptotic properties of
the solutions are also transparently changed, so that the inner horizons of
free static spherically symmetric quintessence lead to a flat spacetime at
infinity ($-\frac{1}{3}<w_q<0$), while the outer horizons produce essentially
non-flat asymptotics, though at typical values expected for the quintessence
energy density the numbers involved might be so small that the effect is
negligible for the real black hole formation until one does not touch the
problem of halos in spiral galaxies.

Finally, Zloshchastiev studied exactly soluble models of
Einstein--Maxwell--dilaton gravity including the case of string-motivated
prescriptions with the exponential potentials \cite{Zl}. His results cover the
situation with both the black hole, which mass can include the logarithmic
variation, and the electric charge varied logarithmically, too, if one suggests
the flat limit of two-dimensional sphere. The horizons and Hawking temperatures
were also studied in \cite{Zl}.

Thus, we see that the perfect fluid under consideration of present work can be
in some sense modelled by a scalar field with an exponential potential,
probably, coupled to a cold dark matter. A motivation for the specific choice
of parameter in the energy-momentum tensor of quintessence by the additivity
and linearity condition may be twofold. First, such the choice matches the
well-known limits as described. Second, it allows us to consistently
incorporate a back reaction of matter on the metrics. In detail, consider the
black hole surrounded by the matter fields. The quantum corrections caused by
the matter loops in the given external classical gravitational field, for
instance, have to influence the metric, too. So, the expectation value of the
energy-momentum tensor for the matter could be considered as a classical source
of gravity. That is the case we consider in the present paper. Hence, if we
suggest validity of the correspondence principle implying the smooth limit of
problem solution with a source to the probelm solution without the source, then
we could expect a continous matching of two exact solutions with the source and
without it. Indeed, imagine we introduce a small amount of spherically
symmetric matter in the field of black hole. One could expect that, by
continuity, the exact solution with matter is close to the exact solution
without the matter, i.e., the black hole is not destroyed by a negligible
amount of matter, including the case of perfect fluid with a negative presure
we study. Then, the matter with the negative presure should form a coherent
perfect fluid state, in which the energy-momentum tensor under the spherical
symmetry is given by the expression with the definite relation between the
state parameter $w$ and the fraction of radial and isotropic spatial
components. With another choice of relation between two spherically symmetric
parts in the energy-momentum tensor of perfect fluid, the black hole metric
will be crutially broken by a small amount of matter as was mentioned in the
Introduction, since, for instance, in the case of purely isotropic strength
tensor one found solutions with no black hole at all. Anyway, if the state
parameter is adiabatically varied with the distance in the field theory
framework, the perfect fluid solution should give a meaningful approximation
for the exact quantum state. This solution could be also an attractor of
fluctuations in the real problem with a scalar field and a cold dark matter,
which is a question for a study elsewhere.

If the quintessence field is varying, this could possibly have an effect in the
past. In this work we do not concern for the cosmological aspects of the static
quintessence considered.

This work is supported in part by the Russian Foundation for Basic Research,
grants 01-02-99315, 01-02-16585, and 00-15-96645.

\end{document}